\documentclass[lettersize,journal,twoside]{IEEEtran}
\usepackage{amsmath,amsfonts}
\usepackage{amsmath,amssymb,bm,booktabs,multirow,url}
\usepackage{algorithmic}
\usepackage{algorithm}
\usepackage{array}
\usepackage[caption=false,font=footnotesize,labelfont=rm,textfont=rm]{subfig}
\usepackage{textcomp}
\usepackage{stfloats}
\usepackage{url}
\usepackage{verbatim}
\usepackage{graphicx}
\usepackage{cite}
\usepackage{hyperref}
\usepackage{utfsym}
\usepackage{threeparttable}
\usepackage[usestackEOL]{stackengine}
\begin{document}

\title{The Gap Between Principle and Practice\\ of Lossy Image Coding}

\author{Haotian Zhang and Dong Liu
\thanks{Date of current version \today. 
The authors are with the MOE Key Laboratory of Brain-Inspired Intelligent Perception and Cognition, University of Science and Technology of China, Hefei 230093, China (e-mail: zhanghaotian@mail.ustc.edu.cn; dongeliu@ustc.edu.cn).}}

\maketitle

\begin{abstract}
Lossy image coding is the art of computing that is principally bounded by the image's rate-distortion function. This bound, though never accurately characterized, has been approached practically via deep learning technologies in recent years. Indeed, learned image coding schemes allow direct optimization of the joint rate-distortion cost, thereby outperforming the handcrafted image coding schemes by a large margin. Still, it is observed that there is room for further improvement in the rate-distortion performance of learned image coding. In this article, we identify the gap between the ideal rate-distortion function forecasted by Shannon's information theory and the empirical rate-distortion function achieved by the state-of-the-art learned image coding schemes, revealing that the gap is incurred by five different effects: modeling effect, approximation effect, amortization effect, digitization effect, and asymptotic effect. We design simulations and experiments to quantitively evaluate the last three effects, which demonstrates the high potential of future lossy image coding technologies.
\end{abstract}





\section{Introduction}
Data compression, known as source coding in Shannon's information theory, aims to represent data into fewer symbols. Lossless source coding compresses discrete data without any loss of information. In contrast, lossy source coding achieves higher compression ratio by allowing for some loss of information, focusing on removing less important information. Due to limitations in storage and transmission capacity, lossy source coding is much more widely adopted in multimedia applications, such as audio, image, and video compression. In recent years, there has been rapid development of lossy source coding, primarily owing to deep learning technologies. However, the extent to how far current schemes are from the limit of compression performance remains insufficiently explored.

The rate-distortion function \cite{shannon1959coding}, defined as the minimum rate required to compress a data source given an expected distortion, establishes the fundamental limit on the compression efficiency of lossy source coding. Theoretically, any rate-distortion pairs above the rate-distortion function are achievable. However, calculating the rate-distortion function for real-world data sources with non-analytical distributions, such as natural images, poses significant challenges. Recent studies \cite{balle2018variational} have linked the Lagrangian-relaxed rate-distortion function to the optimization objectives of variational auto-encoders \cite{kingma2013vae}. Subsequent studies \cite{yang2021estimterd, duan2023improverd} have introduced upper-bound algorithms for estimating the rate-distortion function of generic sources, implying substantial potential for enhancing the performance of practical lossy source coding.

Moreover, the rate-distortion theory \cite{rdtheory} does not provide a direct method for designing lossy source coding schemes. Such design remains an art. It is true that the rate-distortion theory provides the guidance to design and optimize lossy source coding schemes, which aim to approach the rate-distortion function even though its exact form is still unknown. It is equally true that certain relaxations and assumptions are introduced during the design and optimization processes, incurring a gap between the principle and the practice of lossy source coding.

This paper aims to investigate the gap between the ideal performance forecasted by Shannon's information theory and the empirical performance achieved by lossy source coding schemes, with a focus on lossy image coding.
The state-of-the-art lossy image coding schemes usually adopt a latent-based computing model, i.e., an image is converted to a set of features (aka latents) via the analysis process, and then the latents are entropy encoded into symbols; the symbols are entropy decoded to the latents, and then the latents are converted back to a reconstructed image via the synthesis process \cite{langdon1984ae,flamich2020compressing}.
For this paradigm, we identify that the principle-practice gap is incurred by five different effects:
\begin{itemize}
    \item \textbf{Modeling effect} comes from the model itself, specifically the entropy model used for the latents as well as the synthesis process. 
    \item \textbf{Approximation effect} comes from the restriction of the adopted computing architecture, i.e., the theoretically optimal computing model may not be found.
    \item \textbf{Amortization effect} refers to the difference between optimizing the model parameters over the entire training set and optimizing the model parameters for each sample individually.
    \item \textbf{Digitization effect} refers to the difference between analog communications and digital communications, where the latter is assumed by the state-of-the-art lossy image coding. Digitization causes difficulty in obtaining the optimal model parameters.
    \item \textbf{Asymptotic effect} reflects the fact that the theoretical rate is achievable in the asymptotic setting but practical schemes deal with finite-length images.
\end{itemize} 
Following the analyses, we design simulations and experiments to quantitively evaluate the last three effects. First, on top of a model trained by a large image set, we turn on per-sample optimization to further enhance the compression efficiency, which quantifies (at least) how large the gap is due to the amortization effect. Second, we estimate the theoretical achievable rate (as the mutual information) of an analog communication system and compare with the actual coding rate of the corresponding digital communication system, which quantifies how large the gap is due to the digitization and asymptotic effects. Our results demonstrate that the estimated rate-distortion function achieves 35\% rate savings than the H.266/VVC reference software--VTM on the Kodak image set, which sets up a new benchmark for lossy image coding. Accordingly, it reveals the high potential of future lossy image coding technologies.

\section{Related Work}
\subsection{Practices of Lossy Image Coding}
Lossy image coding is a fundamental problem in information theory. Over the past several decades, extensive research has focused on designing and optimizing practical lossy image coding schemes. In the following, we introduce the deterministic and stochastic practices separately.

\subsubsection{Deterministic Schemes}
Most approaches rely on constructing deterministic systems that heavily rely on quantization to discard information from the source image data. A common framework followed by many methods is the transform coding scheme \cite{goyal2001transform}, where images are first transformed into a latent space to achieve decorrelation, followed by quantization and entropy coding. 

In traditional lossy image codecs such as JPEG \cite{wallace1991jpeg}, JPEG2000 \cite{skodras2001jpeg}, BPG \cite{Bellard2015BPG}, and VVC \cite{bross2021overview}, the different modules—transform, quantization, and entropy coding—are manually designed and optimized separately, rather than in an end-to-end manner.

In recent years, learned lossy image coding methods have achieved superior performance by leveraging the capabilities of neural networks. Unlike traditional approaches, learned lossy image coding \cite{balle2016end, balle2017end} optimizes the various components in an end-to-end manner. 
While some methods \cite{agustsson2017soft, zhu2022unified, feng2023nvtc, zhang2023lvqac} have explored vector quantization for end-to-end optimization, most studies adopted uniform scalar quantization, which currently represents the state-of-the-art approach in learned lossy image coding. Therefore, in the following, we focus specifically on methods based on uniform scalar quantization.

In the typical framework of learned image coding \cite{balle2017end}, the encoder applies an analysis transform to an input image $x$, producing a latent representation $g_a(x|\phi)$. The latent representation is then quantized to obtain the discrete latent $y = \lfloor g_a(x|\phi) \rceil$. This discrete latent is encoded losslessly using an entropy model $Q_{Y}(y|\psi)$. Finally, the decoder recovers $y$ and reconstructs the image as $\hat{x} = g_s(y|\theta)$ using the synthesis transform. The parameters $\phi$, $\theta$, and $\psi$ represent the trainable variables for the analysis transform, synthesis transform, and entropy model, respectively.

Lots of researchers have contributed to the design and optimization of learned lossy image coding methods, which can be primarily categorized into three types: the design of neural network-based transforms, the development of entropy models, and end-to-end optimization techniques. 

The effectiveness of transforms plays a crucial role in approximating the optimal mapping from the source space to the latent space. Recurrent neural networks have been employed in several studies \cite{toderici2015variable, toderici2017full, johnston2018improved, lin2020spatial}. Balle \textit{et al.} \cite{balle2016end} introduced a convolutional neural network-based model for image compression. Additionally, more powerful convolutional neural networks, such as the residual connection and the attention mechanism, are adopted in the following studies \cite{chen2021end, cheng2020learned, guo2021causal, gao2021neural, zou2022devil}. Some researchers \cite{zhu2022transformer, zou2022devil, liu2023learned, li2023frequency} constructed transformer-based transforms. Additionally, instead of using non-invertible transforms, some studies utilized invertible transforms \cite{ma2020end, helminger2020lossy, xie2021enhanced, dong2024wavelet}. 

Entropy models are essential for estimating the distribution of latent variables. A factorized prior was employed by \cite{balle2016end}. Balle \textit{et al.} \cite{balle2018variational} then proposed a hyperprior model that parameterizes the distribution as a Gaussian model conditioned on side information. Minnen \textit{et al.} \cite{minnen2018joint} introduced a more accurate entropy model that jointly utilizes a context model and hyperprior. Numerous studies \cite{lee2018context, chen2021end, cheng2020learned, minnen2020channel, guo2021causal, hu2021learning, qian2022entroformer, he2022elic, mentzer2023m2t, jiang2023mlic, li2023flexible, fu2023learned} have contributed to enhancing entropy models for improved performance.

Since the gradient of the quantization operator is zero almost everywhere, standard back-propagation is inapplicable during training. Various quantization surrogates have been proposed to enable end-to-end optimization. Training with additive uniform noise \cite{balle2016end} is a widely used method for approximating rounding. In \cite{theis2017lossy}, the straight-through estimator \cite{bengio2013estimating} was adopted for training, applying stochastic rounding in the forward pass while utilizing a modified gradient in the backward pass. The study \cite{minnen2020channel} empirically introduced a mixed quantization surrogate, which employs noisy latent for rate estimation but utilizes rounded latent and the straight-through estimator when passing through the synthesis transform. Additionally, the discrepancy in optimization objectives between training and testing, known as the train-test mismatch, has been investigated in studies such as \cite{agustsson2020universally, guo2021soft, yang2020improving, zhang2023uniform}. Some studies \cite{tsubota2023comprehensive, zhang2023uniform} also conducted lots of empirical experiments to evaluate the performance of different quantization surrogates.

\subsubsection{Stochastic Schemes}
Some recent studies have attempted to establish stochastic lossy image coding systems to reduce train-test mismatches caused by quantization surrogates. In these systems, the transmitted latent is represented as $y=g_a(x|\phi)+\epsilon$, where $\epsilon$ is random noise. For instance, using uniform noise instead of rounding during testing is tried in \cite{agustsson2020universally}, but its performance was found to be worse than using rounding \cite{balle2020nonlinear, theis2021advantages}. Other studies \cite{flamich2020compressing} have experimented with Gaussian noise. Although these approaches based on Gaussian noise demonstrated promising theoretical ideal rate-distortion performance, the actual rate implemented may exhibit significant overhead \cite{li2018SFRL}, leading to suboptimal results.

\subsection{Estimating Rate-Distortion Function}
Rate-distortion theory defines the rate-distortion function as the theoretical limit of lossy image coding. However, the calculation of this function for a general source is not specified in rate-distortion theory. Estimating the rate-distortion function for real-world sources is crucial for evaluating the potential improvements in practical lossy image codecs.

The classic Blahut-Arimoto algorithm \cite{Blahut1972ba, Arimoto1972ba} offers an optimization method for calculating the rate-distortion function of a discrete source with a known distribution. The study \cite{harrison2008estimation} theoretically proposed a method for estimating the rate-distortion function based on empirically observed data samples. 
Recent advancements have incorporated deep learning techniques to estimate the rate-distortion function \cite{yang2021estimterd, lei2022nerd, duan2023improverd, yang2024estimating}. Notably, Balle \textit{et al.} \cite{balle2018variational} demonstrated a tight connection between the optimization of $\beta$-Variational Auto-Encoder ($\beta$-VAE) \cite{higgins2017beta} and the Lagrange-relaxed rate-distortion function. Furthermore, several studies \cite{yang2021estimterd, duan2023improverd} have established upper bounds for the rate-distortion function for natural images through training a $\beta$-VAE. Their findings indicate significant potential for enhancing existing lossy image codecs.

\section{Theoretical Analyses}
The aforementioned image coding technologies all belong to the practice of lossy source coding \cite{Berger1998Lossy}. In Shannon's seminal work \cite{shannon1959coding}, lossy source coding was referred to as source coding ``with a fidelity criterion." For lossy source coding, Shannon defined the rate-distortion function as the theoretical limit of coding efficiency. However, rate-distortion theory does not specify how to design a lossy source coding system that achieves this function for a generic source, nor does it provide a method for calculating the rate-distortion function itself.

In practice, we can adhere to the principles of rate-distortion theory to design and optimize lossy source coding systems, with the hope that these systems will approach the rate-distortion function, even though its exact form remains unknown. With this in mind, we begin by reviewing learned image compression technologies, seeking to understand the gap between theoretical principles and practical implementations.

\subsection{Formulation of Optimization Problem}
In this paper, we use uppercase letters, such as \( Y \), to denote random variables, while lowercase letters, such as \( y \), represent specific samples of these random variables, without distinguishing between continuous and discrete variables. We represent the source by a random variable \( X \) that follows a distribution \( P_X \), with the source alphabet denoted as \( \mathcal{X} \).
Similarly, we define \( \hat{X} \) as the reconstruction, characterized by its distribution \( P_{\hat{X}} \) and alphabet \( \hat{\mathcal{X}} \). Furthermore, we assume that a distortion measure \( \Delta: \mathcal{X} \times \hat{\mathcal{X}} \rightarrow \mathcal{R}_{\geq 0} \) has been established, such as the sum-of-squared-error defined by \( \Delta(X, \hat{X}) = \| X - \hat{X} \|^2 \). The notation \( P_{\hat{X}|X} \) denotes a conditional distribution, which can also be interpreted as a virtual channel between \( X \) and \( \hat{X} \).

The rate-distortion function \( R(D) \) for the given source \( X \) and distortion measure \( \Delta(\cdot,\cdot) \) is defined as
\begin{equation}
    \begin{split}
        R(D)=&\inf_{P_{\hat{X}|X}} I(X;\hat{X}),\\
        &\mbox{subject to } \mathbb{E}_{P_X P_{\hat{X}|X}}[\Delta(X,\hat{X})]\leq D,
    \end{split}
\label{eq_basic_rd}
\end{equation}
where \( I(X;\hat{X}) \) is the mutual information between \( X \) and \( \hat{X} \), and \( \mathbb{E}[\cdot] \) represents the mathematical expectation. The minimization of the mutual information is performed over all feasible \( P_{\hat{X}|X} \) under the distortion constraint. The rate-distortion function is determined by the source distribution \( P_X \) and the pre-defined distortion measure \( \Delta(\cdot,\cdot) \). This function describes the minimum achievable rate under a given distortion threshold, indicating that for any practical compression system, the rate cannot be less than \( R(D) \) if the expected distortion is not larger than \( D \).

Rate-distortion theory cannot be directly applied to evaluate the performance limits of lossy image coding or to design a practical codec. Due to various constraints like computational cost, some relaxations and assumptions are often introduced to solve Eq. (\ref{eq_basic_rd}) and develop lossy image codecs. These limitations contribute to a performance gap between the theoretical foundation and practical implementations of lossy image coding. By reviewing techniques in learned lossy image coding, we attribute this gap to five effects: modeling effect, approximation effect, amortization effect, digitization effect, and asymptotic effect, each of which will be introduced and discussed in the following.

\subsubsection{Lagrangian Relaxation} 
\label{sec:Lagrangian}
Since the rate-distortion function is convex, it can be reformulated as an unconstrained optimization problem to obtain the optimal solution. For any $\lambda>0$, the unconstrained optimization objective is defined as 
\begin{equation}
L (P_{\hat{X}|X}) = I(X;\hat{X}) + \lambda \mathbb{E}_{P_X P_{\hat{X}|X}}[\Delta(X,\hat{X})].
\label{eq_lagrange_rd}
\end{equation}
For a given $\lambda$, a Pareto front of rate-distortion pairs can be achieved by minimizing Eq. (\ref{eq_lagrange_rd}). Each point on rate-distortion function curves can be reached by optimizing Eq. (\ref{eq_lagrange_rd}) through varying $\lambda$. 
However, if we consider optimizing another operational rate-distortion function problem, where constraints exist on the feasible region of $P_{\hat{X}|X}$, the operational rate-distortion function is not necessarily convex. In this scenario, Lagrangian methods can not reach some points on rate-distortion curves \cite{wagner2021neural,yang2023introduction}. However, we can minimize Eq. (\ref{eq_lagrange_rd}) to find the convex hull of the operational rate-distortion function.

\begin{figure}
    \centering
  \subfloat[]
    {  
        \includegraphics[width=0.27\linewidth]{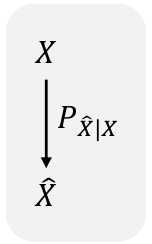}
        \label{fig:P_org}
    } 
    \subfloat[]
    {  
        \includegraphics[width=0.48\linewidth]{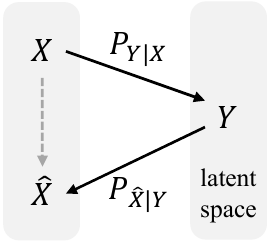}
        \label{fig:P_latent}
    } 
    \caption{(a) Illustration of a virtual channel $P_{\hat{X}|X}$ for lossy source coding; $X$ is the source and $\hat{X}$ is the reconstruction. (b) Illustration of instantiating $P_{\hat{X}|X}$ by a latent-based model; $Y$ refers to the latents.
    }
    \label{fig:P}
\end{figure}

\subsubsection{Modeling Effect} 
\label{sec:modeling}
The rate-distortion function can be estimated using the Monte Carlo method, which involves sampling from the data source,
\begin{equation}
\begin{split}
   I(X;\hat{X}) =& \mathbb{E}_{x\sim P_X}[KL(P_{\hat{X}|X=x}\|P_{\hat{X}})]\\
  \mathbb{E}_{P_X P_{\hat{X}|X}}[\Delta(X,\hat{X})]= & \mathbb{E}_{x\sim P_X}\mathbb{E}_{P_{\hat{X}|X=x}}[\Delta(x,\hat{X})],
\end{split}
\label{eq_mc_rd}
\end{equation}
where $KL(P_{\hat{X}|X=x}\|P_{\hat{X}})$ is the Kullback-Leibler divergence (relative entropy) between $P_{\hat{X}|X=x}$ and $P_{\hat{X}}$. 
Since $P_{\hat{X}}$ requires knowledge of the source distribution, which in practice is usually unavailable. We need another probability measure $Q_{\hat{X}}$ to model the distribution of the reconstruction.
It can be costly when dealing with high-dimensional sources with memory like natural images. Instead of directly modeling the distribution in the source or reconstruction space, practical lossy image coding methods usually adopt a latent variable model as shown in Fig. \ref{fig:P_latent} for simplicity. In this framework, we need to map the source space into the latent space through $P_{Y|X}$, model the distribution of latent variable $Y$ by $Q_{Y}$, and use another conditional distribution $P_{\hat{X}|Y}$ to synthesize reconstruction. In this framework, we have the objective as
\begin{equation}
\begin{split}
   L(P_{Y|X},Q_{Y},P_{\hat{X}|Y})=&I(X;Y)+KL(P_{Y}\|Q_{Y})\\
   &+\lambda\mathbb{E}_{P_X P_{Y|X} P_{\hat{X}|Y}}[\Delta(X,\hat{X})],\\
   I(X;Y)+KL(P_{Y}\|Q_{Y})=&\mathbb{E}_{x\sim P_X}[KL(P_{Y|X=x}\|Q_{Y})].
\end{split}
\label{eq_transform}
\end{equation}
Equation \ref{eq_transform} is minimized it over all feasible $P_{Y|X},Q_{Y},P_{\hat{X}|Y}$. 

Some studies \cite{agustsson2019gancom,ma2021end,yang2023cdc} introduced stochastic elements when synthesizing reconstruction for perception optimization.
We consider the common scenario in learned image compression for distortion optimization, where $P_{\hat{X}|Y}$ is a deterministic transform, denoted as $F_D$. Then the objective becomes
\begin{equation}
\begin{split}
   L(P_{Y|X},Q_{Y},F_D) =& I(X;Y)+KL(P_{Y}\|Q_{Y})\\
   &+ \lambda\mathbb{E}_{P_X P_{Y|X}}[\Delta(X,F_D(Y))].
\end{split}
\label{eq_bijective_dec}
\end{equation}

Assuming we can obtain the optimal $P_{Y|X}$, we define the gap caused by the suboptimality of $Q_{Y}, F_D$ as the modeling effect. The modeling effect consists of two components: the entropy modeling effect and the information loss effect:
\begin{itemize}
    \item \textbf{Entropy modeling effect} comes from the inaccuracies in the chosen $Q_Y$ when approximating $P_Y$, quantified as $KL(P_{Y}\|Q_{Y})$. The gap is closed if $P_Y\equiv Q_Y$ such that $KL(P_{Y}\|Q_{Y})=0$.
    \item \textbf{Information loss effect} results from the information loss during the synthesis process of reconstruction. Since the coding process is a Markov Chain, $X\rightarrow Y\rightarrow \hat{X}$, we have $I(X;Y)\geq I(X;\hat{X})$, making the gap equal to $I(X;Y)-I(X;\hat{X})$. The gap is closed if $F_D$ is bijective such that $I(X;Y)= I(X;\hat{X})$.
\end{itemize}
Many studies \cite{balle2018variational,minnen2018joint,lee2018context, chen2021end, cheng2020learned, minnen2020channel, guo2021causal, hu2021learning, qian2022entroformer, he2022elic,liu2023learned, mentzer2023m2t, jiang2023mlic, li2023flexible, fu2023learned, li2023frequency} focused on designing more effective entropy models to reduce the entropy modeling effect. Some studies \cite{ma2020end, helminger2020lossy} explored invertible neural networks to eliminate the information loss effect.

\subsubsection{Approximation Effect} 
In this section, we focus on the optimization of $P_{Y|X}$ within the system described by Eq. (\ref{eq_bijective_dec})
We denote the optimal $P_{Y|X}$ as $P^*_{Y|X}$. In practice, a specific form of $P_{Y|X}$ is typically assumed when optimizing Eq. (\ref{eq_bijective_dec}). Previous studies usually use a basic form to approximate $P^*_{Y|X}$, such as factorized Gaussian or scalar quantization, primarily due to considerations of computational cost and sampling efficiency. These assumptions restrict the feasible region of $P_{Y|X}$, potentially leading to a suboptimal solution. We denote the restricted feasible solution set as $A$, and the optimal $P_{Y|X}$ within this subset as $P^{(A)*}_{Y|X}$.
The performance gap between $P^*_{Y|X}$ and $P^{(A)*}_{Y|X}$ is referred to as the approximation effect \cite{Cremer2018suboptimal}.

For instance, most studies in learned lossy image coding treated $P_{Y|X}$ as a deterministic mapping that transforms the source space into a discrete latent space. We denote this deterministic mapping as $F_E$. Notably, unlike $F_D$, which can be a bijective transform, $F_E$ cannot be bijective because it is designed to induce information loss to achieve lossy coding. In such a deterministic system, the objective becomes
\begin{equation}
\begin{split}
   L(F_{E},Q_{Y},F_{D}) =& H(Y)+KL(P_{Y}\|Q_{Y}) \\
   &+ \lambda\mathbb{E}_{P_X}[\Delta(X,F_{D}(F_E(X)))],\\
   H(Y)+KL(P_{Y}\|Q_{Y}) =& \mathbb{E}_{P_X}[-\log Q_Y(F_E(X))].
\end{split}
\label{eq_deterministic}
\end{equation}
Some studies treated $P_{Y|X}$ as a conditional uniform distribution \cite{agustsson2020universally}, but its performance was found to be worse than using rounding \cite{balle2020nonlinear, theis2021advantages}
Some studies \cite{flamich2020compressing,yang2021estimterd,duan2023improverd} assumed $P_{Y|X}$ as a conditional Gaussian distribution, which shows the advancement for optimizing Eq. (\ref{eq_bijective_dec}) compared to deterministic systems.

As discussed in Sec. \ref{sec:Lagrangian}, constraints on the feasible region of $P_{Y|X}$ may make the original optimization problem non-convex, such that optimizing the Lagrangian relaxation can only reach a subset in rate-distortion curves. 
Previous studies \cite{gyorgy2000uniform,wagner2021neural} presented some toy examples where the operational rate-distortion function is not convex when \( Y \) is restricted to be discrete. For these sources, deterministic mappings can only address certain target rates.

\subsubsection{Amortization Effect} 
In this section, we explore the implementation of \( P_{Y|X} \). To optimize Eq. (\ref{eq_bijective_dec}), \( P_{Y|X} \) is typically parameterized using neural networks, which are shared across the entire dataset, rather than directly optimizing $Y$ for each sample $x$, which can be costly. For example, many studies parameterized the deterministic mapping described by Eq. (\ref{eq_deterministic}) with scalar quantization as \( Y = F_E(X) = \lfloor g_a(X|\phi) \rceil \), where \( \phi \) represents the parameters that can be optimized. Some other studies parameterized the stochastic \( P_{Y|X} \) as \( Y = g_a(X|\phi) + \epsilon \), with \( \epsilon \) denoting random noise. Theoretically, we could use a sufficiently large number of parameters to fit \( P^{(A)*}_{Y|X} \) within our restricted set \( A \). However, in practice, due to the limitations of neural networks and computational costs, \( P^{(A)*}_{Y|X} \) may not be perfectly implemented. We refer to the performance gap between the implemented \( P^{(A)}_{Y|X} \) and \( P^{(A)*}_{Y|X} \) as the amortization effect \cite{Cremer2018suboptimal}. The term ``amortization" highlights the difference arising from amortizing the parameters \( \phi \) across the entire set of source samples rather than optimizing for each sample individually. 

Many studies \cite{chen2021end, cheng2020learned, guo2021causal, gao2021neural, zou2022devil, zhu2022transformer, zou2022devil, liu2023learned, li2023frequency} attempted to adopt more powerful networks to fit \( P^{(A)*}_{Y|X} \) to reduce this effect. Some studies \cite{schafer2021rdodic,wang2021ensemble} employed an ensemble way to enhance the ability of networks. There is another way \cite{kim2018semi,Marino2018imi,campos2019content,yang2020improving} which further conducts per-sample optimization based on a well-optimized \( P^{(A)}_{Y|X} \) to reduce the amortization effect. For $x\sim P_X$, we can optimize $Y$ through 
\begin{equation}
\begin{split}
Y=\arg\min_{Y}& KL(P_{Y|X=x}\|Q_{Y})\\
&+\lambda\mathbb{E}_{P_{Y|X=x}}[\Delta(x,F_D(Y))].
\end{split}
\label{eq_online_e}
\end{equation}
For a deterministic system described by Eq. (\ref{eq_deterministic}), the objective can be simplified as
\begin{equation}
\begin{split}
y=\arg\min_{y}& -\log Q_Y(y)+\lambda \Delta(x,F_D(y)).
\end{split}
\label{eq_online_e_de}
\end{equation}

\subsubsection{Digitization Effect}
For the systems established in the previous section, a critical challenge lies in the optimization techniques. With advancements in deep learning, gradient descent has become a widely adopted method for optimizing parameters. For continuous variables \( Y \), unbiased path derivative gradient estimation can be efficiently implemented using the reparameterization trick \cite{kingma2013vae}. However, for discrete variables \( Y \), standard gradient descent is not applicable, because the derivative of the quantization operator is zero almost everywhere. The score function-based gradient estimator, also known as REINFORCE \cite{william1992reinforce}, provides an unbiased way to estimate the gradient for discrete \( Y \). Nevertheless, due to the high latent dimensionality of natural images, this approach suffers from excessive variance. As a result, current implementations often rely on path derivative estimators.

We consider the deterministic system described by Eq. (\ref{eq_deterministic}). There are two primary approaches to facilitate optimization using gradient descent. The first approach involves using a differentiable surrogate to replace the quantization operator, allowing for the application of standard gradient descent. However, this leads to an objective mismatch between the training and testing phases. The second approach is to define a derivative for the quantization operator, such as the identity function \cite{bengio2013estimating}, which introduces gradient estimation error. This bias in gradient estimation cannot guarantee convergence of the objective function. The train-test mismatch and gradient estimation error during training for deterministic systems cause the digitization effect.

Uniform scalar quantization, with rounding as its common case, is widely adopted in learned image compression. To enable end-to-end optimization, various quantization surrogates have been proposed to replace rounding. Training with additive uniform noise \cite{balle2016end} is a popular approach for approximating rounding. This method introduces stochasticity during training, which leads to a discrepancy between the training and testing phases. In \cite{theis2017lossy}, the straight-through estimator \cite{bengio2013estimating} is used, where stochastic rounding is applied during the forward pass, and a modified gradient is used in the backward pass. Several studies \cite{minnen2020channel,guo2021soft,tsubota2023comprehensive} have also tried to employ rounding in the forward pass and the straight-through estimator in the backward pass, avoiding train-test mismatch but introducing significant gradient estimation errors, which results in poor performance.
Minnen and Singh \cite{minnen2020channel} empirically proposed a mixed quantization surrogate, which uses noisy latent for rate estimation but applies rounded latent and the straight-through estimator during reconstruction synthesis. This mixed surrogate has demonstrated superior performance in several learned image compression models and has been widely adopted in recent state-of-the-art methods \cite{he2022elic,liu2023learned,jiang2023mlic,li2023frequency}. 

Due to the train-test mismatch inherent in the mixed quantization surrogate, optimizing the surrogate objective, denoted as \( \widetilde{L} \), does not necessarily guarantee the minimization of the true objective \( L \). Additionally, gradient estimation errors can lead to suboptimal solutions for \( \tilde{L} \). As a result, the performance of the obtained solution \((F_{E}, Q_{Y}, F_{D})\) may be significantly worse than that of the optimal solution for \( L \).

The digitization effect also arises in per-sample optimization described in Eq. (\ref{eq_online_e_de}). The study \cite{yang2020improving} suggested employing an annealing proxy during per-sample optimization to reduce the digitization effect. By gradually decreasing the temperature coefficient, the proxy objective 
\( \tilde{L} \) converges toward the true objective \( L \).

\subsubsection{Asymptotic Effect}
The achievability of the rate-distortion function is established within the asymptotic setting, while the one-shot setting is standard in practical usage, thereby creating an asymptotic effect. In the asymptotic scenario, the average rate can approach mutual information as the dimension of the expanded source tends to infinity. In contrast, in the one-shot setting—where each sample is compressed individually—the rate may not be able to reach this lower bound.

In the systems described in Eq. (\ref{eq_bijective_dec}), without any assumptions regarding the distributions of \( X \) and \( Y \), previous studies \cite{li2018SFRL,li2021PML} have proved a bound on the achievable rate in the one-shot setting, which is given by
\begin{equation}
\begin{split}
I \leq & R \leq I + \log(I + 1) + O(1),
\end{split}
\label{lKL}
\end{equation}
where $I = I(X;Y)+KL(P_{Y}\|Q_{Y})$ and $O(1)$ represents a constant. The overhead $\log(I + 1) + O(1)$ can be relatively small if the mutual information is large. However, it may be computationally expensive to communicate more information at once \cite{agustsson2020universally} in practice.
For some specific distributions, this bound can be significantly tightened. For instance, if \( P_{Y|X} \) is a conditional factorized uniform distribution, the rate can be achieved through universal quantization \cite{ziv1985universal,zamir1992universal}, which is bounded by
$I \leq R \leq I + O(1)$, where $I = I(X;Y)+KL(P_{Y}\|Q_{Y})$ .

For a deterministic system described by Eq. (\ref{eq_deterministic}), we only need to losslessly encode the discrete variable $Y$. The achievable rate in the one-shot setting satisfies \( H \leq R \leq H + O(1) \), where $H = H(Y)+KL(P_{Y}\|Q_{Y})$.

Some studies \cite{Theis2022rcc,flamich2020compressing} have attempted to establish universal algorithms for communicating samples. These methods have also been investigated within the context of learned image compression \cite{flamich2020compressing}. Modeling \( P_{Y|X} \) as a conditional Gaussian yields impressive results in estimating rate-distortion functions; however, performance in the one-shot setting remains inferior to that of deterministic systems. The overhead of the rate caused by the asymptotic effect is substantial for such a system. 

\subsection{Advanced Learned Image Coding Methods}
Learned lossy image coding methods have demonstrated significant performance improvements compared to traditional lossy image codecs. Among these methods, deterministic systems described in Eq. (\ref{eq_deterministic}), have achieved state-of-the-art compression performance \cite{liu2023learned,jiang2023mlic,li2023frequency}. The study \cite{theis2021advantages} theoretically highlights the superiority of deterministic systems in the one-shot setting. In this section, we examine the gap between this scheme and the theoretical limit.

We begin by analyzing how each module in the scheme contributes to the gap, including the analysis transform \( g_{a} \), the synthesis transform \( g_s \), the entropy model \( Q_Y \), the quantization surrogate used during training (e.g., additive uniform noise), and the entropy coding engine (e.g., arithmetic coding). In this scheme, we have \( F_{E}(x) = \lfloor g_a(x) \rceil \) and \( F_{D}(y) = g_s(y) \). The quantization surrogate is employed during training for end-to-end optimization. In the encoding process, the analysis transform is applied to a source image $x$, producing a latent representation $g_a(x)$. The latent representation is then quantized to obtain the discrete latent $y = \lfloor g_a(x) \rceil$. This discrete latent is losslessly encoded into bitstreams by the entropy coding engine, utilizing the entropy model $Q_{Y}(y)$. In the decoding process, $y$ is first losslessly recovered and then used to generate the reconstruction as $\hat{x} = g_s(y)$ through the synthesis transform. 

The analysis transform primarily contributes to the amortization effect, digitization effect, and modeling effect. For a given \( g_s \) and \( Q_Y \), the amortization effect is determined by the analysis transform. Employing a more powerful neural network as the analysis transform can reduce this gap. Additionally, per-sample optimization techniques, such as those proposed in \cite{campos2019content,yang2020improving}, can enhance the generalized analysis transform to further mitigate the amortization effect.
The decorrelation capability of the analysis transform influences the modeling effect, since the optimization of the entropy model relies on the analysis transform. A more powerful analysis transform can alleviate the burden on the entropy model, thereby reducing the entropy modeling effect.
The digitization effect is inevitable in systems that rely on quantization, particularly during the optimization for the analysis transform. Due to the presence of train-test mismatch and gradient estimation error, a more powerful analysis transform may not consistently reduce the amortization effect. In fact, employing a stronger analysis transform could potentially increase the digitization effect. The study \cite{zhang2023uniform} has shown some cases where utilizing more powerful networks as the analysis transform leads to degraded compression performance with certain quantization surrogates.

The synthesis transform primarily contributes to the modeling effect, amortization effect, and digitization effect. The function of the synthesis transform is to generate reconstructions from the latent variables and the optimal synthesis process is determined by the analysis transform. As discussed in Sec. \ref{sec:modeling}, the modeling effect introduced by the synthesis transform could be eliminated by using an invertible network. In practice, due to the limited capabilities of the analysis transform and entropy model, the optimal synthesis process may take on a complex form. Employing a more powerful network as the synthesis transform can alleviate the burden on the analysis transform and the entropy model, thereby impacting both the amortization effect and the entropy modeling effect.
Empirical results \cite{xie2021enhanced,ma2020end,dong2024wavelet} demonstrate that using current invertible neural networks as synthesis transforms leads to suboptimal performance in practice. Consequently, researchers have opted for more powerful non-invertible neural networks to enhance overall performance.
Regarding the digitization effect, the optimization of the synthesis transform is affected by the train-test mismatch. The extent of the influence of train-test mismatch is closely related to the characteristics of the synthesis transform \cite{zhang2023uniform}.

The entropy model influences the modeling effect, amortization effect, and digitization effect. Enhancements in entropy modeling can reduce the entropy modeling effect through advanced structures like autoregressive models \cite{minnen2018joint}, latent variable models \cite{balle2018variational}, flexible probabilistic models \cite{balle2016end,cheng2020learned}, and more powerful neural networks. Since the analysis transform and entropy model are jointly optimized, the ability of the entropy model also influences the amortization effect. Considering the digitization effect, similar to the synthesis transform, the optimization of the entropy model is affected by the train-test mismatch. The extent of the influence is related to the characteristics of the entropy model \cite{zhang2023uniform}.

The quantization surrogate is a significant contributor to the digitization effect. Studies like \cite{zhang2023uniform} and \cite{tsubota2023comprehensive} indicate that the optimal choice of quantization surrogate differs among learned lossy image coding methods. Specifically, \cite{zhang2023uniform} highlights two critical elements influencing the effectiveness of quantization surrogates: train-test mismatch and gradient estimation error. This study indicates a tradeoff exists between the train-test mismatch and the gradient estimation risk, and the optimal balance varies across different network structures. However, how much the digitization effect affects compression performance is still not clearly understood. In per-sample optimization, the digitization effect remains, but it is simplified because the entropy model and synthesis transform are fixed, leaving only the optimization for a single sample. Yang \textit{et al.} \cite{yang2020improving} proposed an annealing-based method to reduce this gap.

The entropy coding engine contributes to the asymptotic effect. Recent advanced methods usually use arithmetic coders \cite{langdon1984ae} or asymmetric numeral systems \cite{duda2013asymmetric} to encode symbols into bitstreams based on the entropy model. Since a large number of symbols are required to represent a natural image, the overhead of coding costs compared to the ideal rate is negligible.

In summary, in the current advanced deterministic schemes, the asymptotic effect is negligible, while the modeling effect, amortization effect, and digitization effect play crucial roles. To address the modeling and amortization effects, we can leverage more advanced networks or structures, such as large-scale generative models. This approach has demonstrated significant performance improvements in learned lossless image coding \cite{li2024understanding}, where the digitization effect is not a concern. However, in lossy image coding, the digitization effect introduces significant uncertainty, complicating performance optimization and potentially limiting the achievable compression performance.

\subsection{Estimating Rate-Distortion Function}
To estimate an upper bound of the rate-distortion function for natural images, we focus only on minimizing Eq. (\ref{eq_bijective_dec}) without the need to establish a practical implementation, without regarding the asymptotic effect. Studies such as \cite{flamich2020compressing, yang2021estimterd, theis2022lossy} demonstrated the superior performance of continuous stochastic systems in optimizing this objective, without regarding the digitization effect. Additionally, a recent study \cite{duan2023improverd} indicates that the estimated upper bound of the rate-distortion function can achieve up to 
30\% rate savings compared to VVC on the Kodak dataset, underscoring the potential of current lossy image coding methods.

When estimating this upper bound through Eq. (\ref{eq_bijective_dec}) using a continuous stochastic system \cite{yang2021estimterd}, we need to address the modeling effect, approximation effect, and amortization effect. As with deterministic systems, the modeling effect and amortization effect can be reduced by utilizing more powerful networks, and per-sample optimization can further mitigate the amortization effect. To address the approximation effect, a more effective distribution can be selected to approximate $P^*_{Y|X}$. Notably, factorized Gaussian distributions have shown impressive results for estimating the upper bound of the rate-distortion function for natural images, while more flexible distributions may further reduce the approximation effect \cite{rezende2015variational}.

\section{Experimental Analyses}
In addition to theoretical analyses, we conduct experimental investigations to gain insights into the magnitude of the performance gap between practice and principle. We employ the upper-bound algorithm in \cite{yang2021estimterd} to estimate the ideal performance predicted by rate-distortion theory. Various structures and complexities are utilized to evaluate both the modeling effect and the amortization effect. By comparing the performance of the upper-bound algorithm with empirical performance in the one-shot setting involving quantization, we aim to provide some insights into the digitization effect.

\subsection{Experimental Setting}
\subsubsection{Learned Image Coding Models}
We employed two structures for our experiments: the mean-scale hyperprior model \cite{minnen2018joint} and a model that incorporates more advanced entropy models \cite{li2023neural}. All transform modules are constructed using partial convolution residual blocks \cite{chen2023pconv}, while depth-wise convolution residual blocks \cite{chollet2017xception} are utilized in context models. We refer to the mean-scale hyperprior model with modified transforms as \textbf{Res-hyper}, and the version that integrates the advanced context model from \cite{li2023neural} as \textbf{Res-context}. We adjust the modeling effect and amortization effect by adjusting the complexity of the neural networks. Specifically, we trained 8 Res-hyper models with various numbers of parameters by gradually increasing the depth and width of residual blocks in transforms. We trained one Res-context model by additionally introducing the context model on the largest Res-hyper model.

\subsubsection{Training} 
We utilized Mean Squared Error (MSE) in the RGB color space as the distortion metric, optimizing all models for this criterion. Multiple models were trained with varying values of \(\lambda \in \{0.0018, 0.0054, 0.0162, 0.0483\}\). The training dataset contains approximately 20000 images from the Flicker2W dataset \cite{liu2020unified}. During each training iteration, images were randomly cropped into $256\times 256$ patches.

For training, we employed the Adam optimizer \cite{kingma2014adam} for 500 epochs, with a batch size of 8 and an initial learning rate of \(5 \times 10^{-5}\). After 400 epochs, the learning rate was reduced to \(2 \times 10^{-5}\), and after an additional 50 epochs, it was further decreased to \(5 \times 10^{-6}\). A consistent random seed was utilized across all experiments.

In training deterministic systems, we trained with mixed quantization surrogates, zero-center quantization \cite{minnen2020channel}, and lower bounds for Gaussian scale parameters \cite{zhang2023uniform} for all models. 

To estimate the upper bound of the rate-distortion function, we employed the Gaussian conditional distribution \cite{yang2021estimterd}. The Gaussian scale parameters were exponentially activated to avoid approaching zero. We calculated the estimated rate as
\begin{equation}
\begin{split}
\mbox{For }& x\sim P_X, P_{Y|X=x}\sim\mathcal{N}(\mu,\sigma), Q_{Y}\sim\mathcal{N}(\hat{\mu},\hat{\sigma})\\
R &= KL(P_{Y|X=x}\|Q_{Y}) \\
&=\left(\log\frac{\hat{\sigma}}{\sigma}+\frac{\sigma^2+(\mu-\hat{\mu})^2}{2\hat{\sigma}^2}-\frac{1}{2}\right)/\log(2) \mbox{ bits}.
\end{split}
\label{eq_bias}
\end{equation}

\begin{figure}[!t]
\centering
  \subfloat[BD-rate of the estimated rate-distortion function over the performance of VTM]
    {  
        \includegraphics[width=0.95\linewidth]{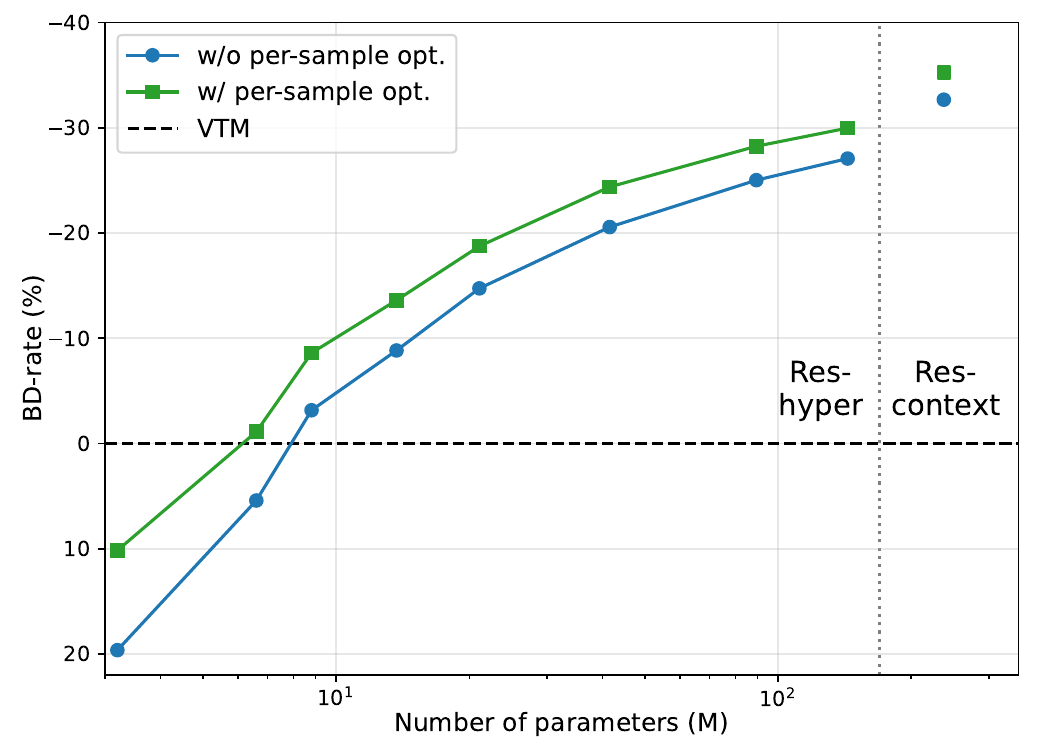}
        \label{fig:bd_para_vae}
    } 
    
    \subfloat[BD-rate of the empirical rate-distortion function over the performance of VTM]
    {  
        \includegraphics[width=0.95\linewidth]{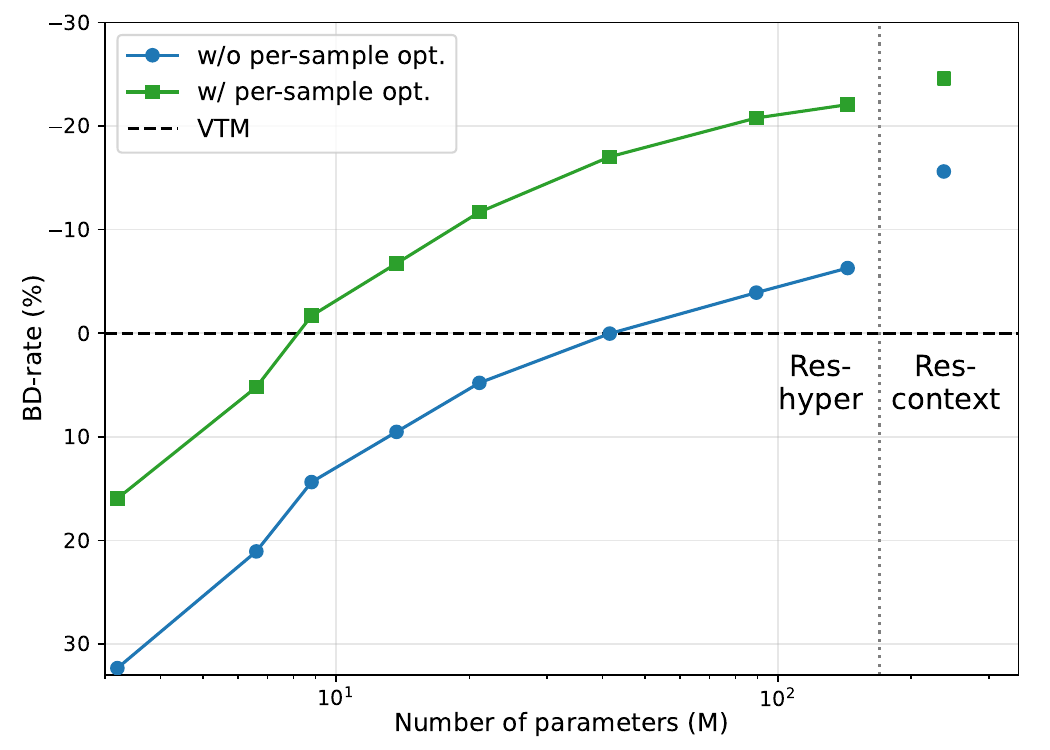}
        \label{fig:bd_para_quant}
    } 
    \caption{The compression efficiency of the estimated and empirical rate-distortion functions with different models and different numbers of parameters. A negative BD-rate indicates rate saving.}
    \label{fig:bd_para}
\end{figure}

\subsubsection{Per-Sample Optimization}
Per-sample optimization was applied to reduce the amortization effect by optimizing all latent variables with the Adam optimizer. There are two latents in our model, the main latent $Y$ and the hyperlatent $Z$, which characterizes the distribution of $Y$. Both $Y$ and $Z$ are optimized together. The initial learning rate was set to \(5 \times 10^{-3}\), and it was gradually annealed using a cosine schedule to \(10^{-4}\) over 5000 iterations. For optimizing deterministic systems, Stochastic Gumbel Annealing (SGA) \cite{yang2020improving} was employed to replace the rounding operator.

\subsubsection{Evaluation}
We evaluate the performance using the Kodak dataset \cite{franzen1999kodak}, which consists of 24 images with resolutions of either 512×768 or 768×512 pixels. To quantify rate and distortion, we use bits-per-pixel (bpp) and Peak Signal-to-Noise Ratio (PSNR), where PSNR is measured in the RGB color space. Additionally, we employ the BD-Rate metric \cite{bjontegaard2001calculation} to compute rate savings. For the overall BD-Rate across the dataset, we calculate the BD-Rate for each image and then average these values. We also test the performance of VTM-22.0\footnote{https://vcgit.hhi.fraunhofer.de/jvet/VVCSoftware\_VTM}, using the YUV444 input format for comparison.

\subsection{Analyses}

\begin{figure}[!t]
    \centering
    \includegraphics[width=0.95\linewidth]{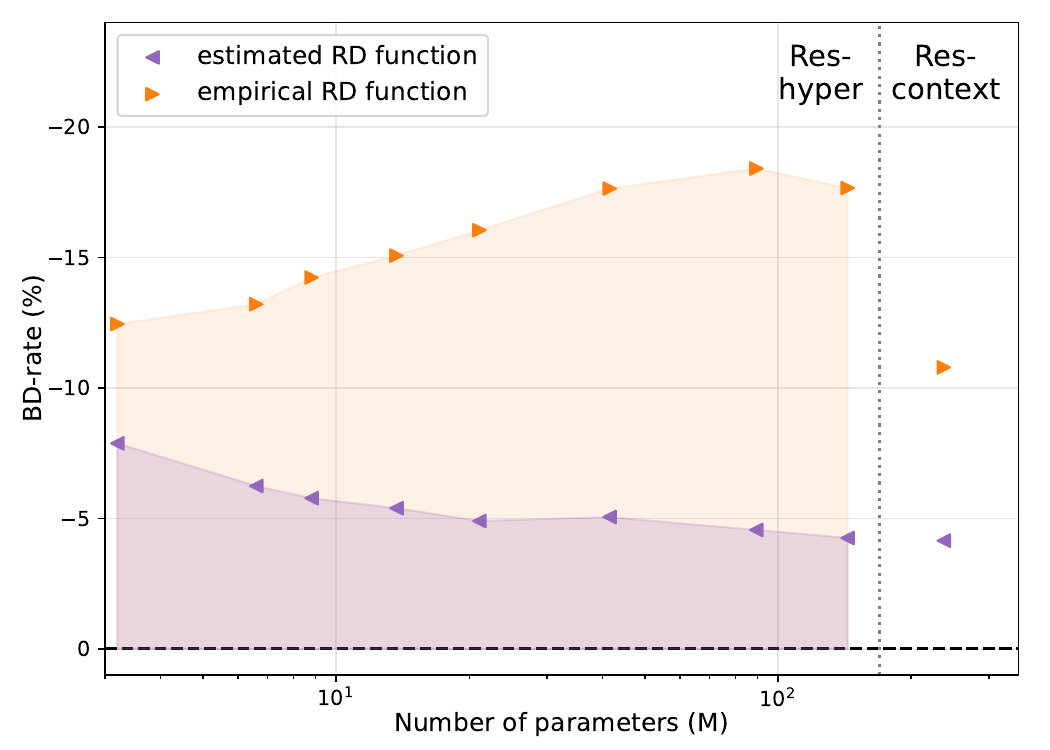}
    \caption{The compression efficiency gain provided by the per-sample optimization on each model.
    }
    \label{fig:bd_para_online}
\end{figure}

\begin{figure}[!t]
    \centering
    \includegraphics[width=0.95\linewidth]{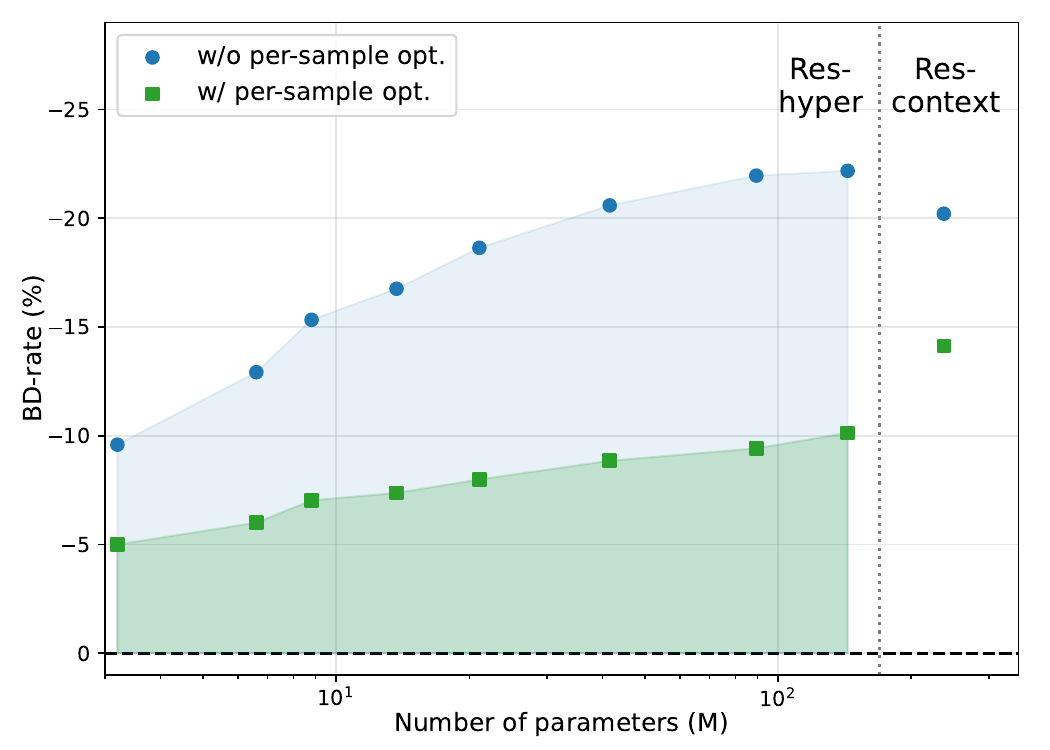}
    \caption{The compression efficiency gap between the estimated and the empirical rate-distortion functions for different models and different numbers of parameters.
    }
    \label{fig:bd_para_vs}
\end{figure}

\begin{figure}[!t]
    \centering
    \includegraphics[width=0.95\linewidth]{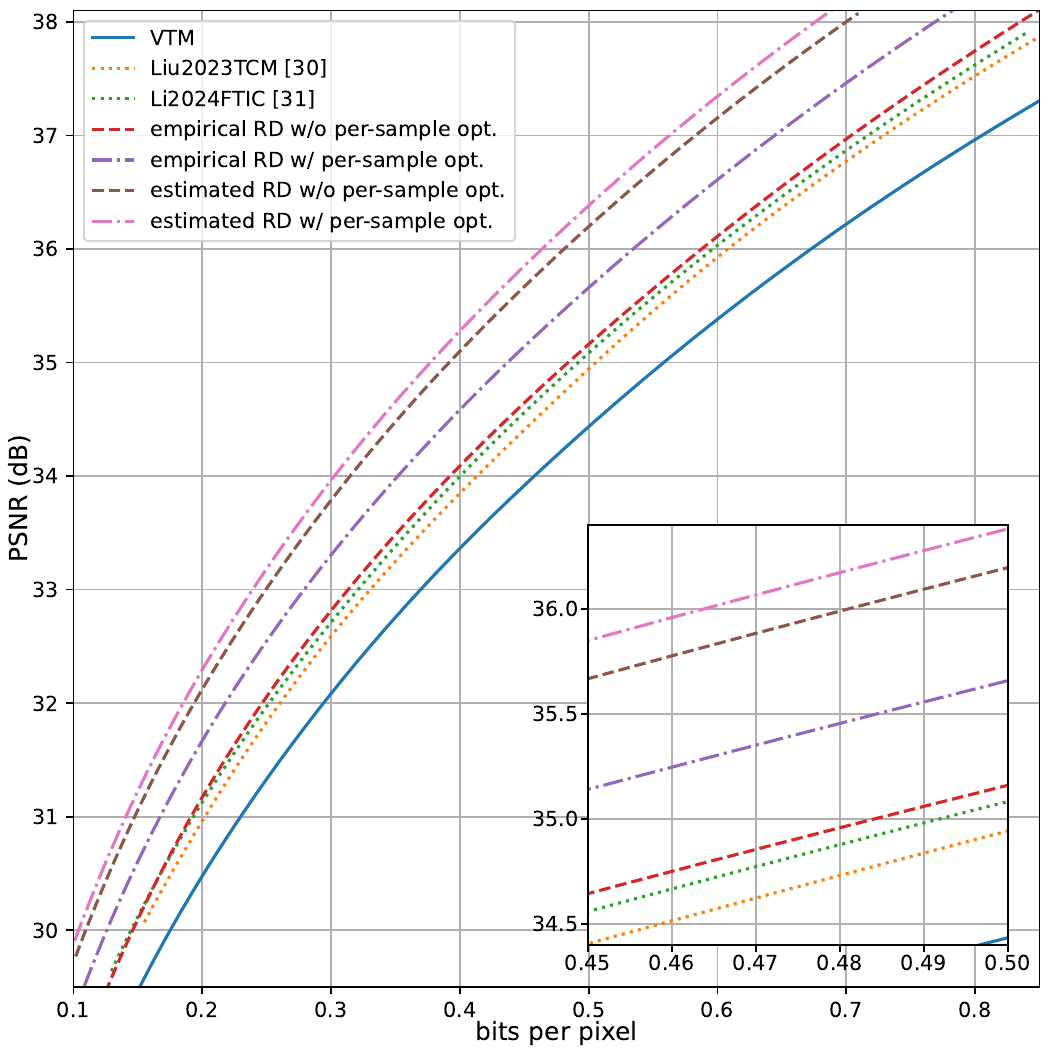}
    \caption{Rate-distortion curves of different methods. The empirical and estimated RD shown in this figure is the performance of the Res-context model.
    }
    \label{fig:rd}
\end{figure}

As illustrated in Fig. \ref{fig:bd_para_vae}, the performance of the estimated bounds by the Res-hyper model improves as the number of parameters in the transform modules increases, although the marginal gains gradually diminish. The Res-context model, which uses a more advanced entropy model, shows better overall performance and larger marginal gains as the number of parameters continues to increase. Per-sample optimization, which enhances the analysis transform, consistently improves performance, but the gains decrease as model complexity increases, as illustrated in Fig. \ref{fig:bd_para_online}. By using more powerful transforms and entropy models, we can tighten the estimated upper bound of the rate-distortion function by reducing both the amortization and entropy modeling effects.

As shown in Fig. \ref{fig:bd_para_quant}, the empirical performance of the Res-hyper model in one-shot setting also improves as the number of parameters increases, but the speed of improvement is slower than that of the estimated bound, as depicted in Fig. \ref{fig:bd_para_vs}. Furthermore, per-sample optimization proves to be more effective for enhancing empirical performance compared to the estimated bound, as illustrated in Fig. \ref{fig:bd_para_online}. Although per-sample optimization with the SGA surrogate is also affected by the digitization effect, it can reduce the gap introduced during training to some extent. These results provide insights into the magnitude of the digitization effect, which can be significant.
The Res-context model shows better practical performance and larger marginal gains as the number of parameters continues to increase. An interesting phenomenon in Fig. \ref{fig:bd_para_vs} reveals that as the number of parameters increases, the gap between the estimated bound and the empirical performance for the Res-hyper model widens, while using the Res-context model results in a smaller gap with an increased number of parameters compared to the Res-hyper model. This suggests that the digitization effect may vary across different models. 

Our experimental results also show that current practical lossy image coding schemes can achieve significant performance improvements by employing more advanced transform modules, entropy models, and per-sample optimization, resulting in up to 24\% rate savings compared to VTM on the Kodak dataset. Furthermore, the estimated upper bound of the rate-distortion function has been tightened to provide up to 35\% rate savings relative to VTM. As shown in Fig. \ref{fig:rd}, the estimated upper bound of the rate-distortion function significantly outperforms the current advanced learned image compression models \cite{liu2023learned,li2023frequency}, which reveals the high potential of future lossy image coding technologies. When evaluating our model using MS-SSIM, the empirical rate-distortion function and the estimated rate-distortion function achieve rate savings of -22\% and -36\% compared to VTM, respectively.

\section{Conclusion}
In this paper, we analyze the gap between the ideal rate-distortion function and the empirical rate-distortion function for lossy image coding. We reveal that the gap is due to five different effects: modeling effect, approximation effect, amortization effect, digitization effect, and asymptotic effect. We analyze each effect by reviewing the technologies currently used in lossy image coding practices. Our simulation and experimental results illustrate how these effects impact compression performance. Additionally, we advance both the estimated and the empirical rate-distortion performance, demonstrating the high potential of future technologies.

\bibliographystyle{IEEEtran}
\bibliography{main}

\vfill
\end{document}